\documentclass[seceq]{ptptex}
\usepackage{type1cm, 
hyperref}
\newcommand{\al}{\alpha'}
\newcommand{\g}{\gamma}

\newcommand{\be}{\begin{equation}}
\newcommand{\ba}{\begin{eqnarray}}
\newcommand{\ea}{\end{eqnarray}}
\newcommand{\ee}{\end{equation}}

\newcommand{\f}{\frac}

\newcommand{\bp}{\bar{\partial}}
\newcommand{\p}{\partial}

\newcommand{\ti}{\tilde}

\title{%        %You can use \\ for explicit line-break
D-branes in the Lorentzian Melvin Geometry}
\author{%       %Use \sc for the family name
Ta-Sheng {\sc Tai}\footnote{E-mail: tai@hep-th.phys.s.u-tokyo.ac.jp}}
\inst{%         %Affiliation, neglected when [addenda] or [errata]
Department of 
Physics, University of Tokyo, Tokyo 113-0033, Japan}

\recdate{%      %Editorial Office will fill in this.
Received October 13, 2006}

\abst{%       %this abstract is neglected when [addenda] or [errata] 
We consider string theory on the Lorentzian Melvin geometry, 
which is obtained by analytically continuing 
the two-parameter Euclidean Melvin background. Because 
this model provides a solvable conformal field theory that describes time-dependent 
twisted string dynamics, 
we study the string one-loop partition function and the D-brane spectrum. 
We found that both the 
wrapping D2-brane and the codimension-one D-string emit winding strings, 
and this behavior can be traced to the modified 
open string Hamiltonian on these probe D-branes.}

\newpage

\begin{document}

\maketitle

\section{Introduction and summary}
In recent years, the string theory approach has 
provided insights into the study of time-dependent 
and cosmological backgrounds. Among the theoretical constructs 
provided by this approach, time-dependent string orbifolds 
are of interest because they are solutions of the string equations 
of motion to all orders in $\al$ and exactly solvable.%
\footnote{There are 
many works studying time-dependent string backgrounds, e.g. 
Refs. 1) - 17). Aspects of string (and D-brane) dynamics 
are explored in Refs. 4), 18) - 27). 
Other issues regarding 
curvature singularity resolution and the 
matrix model are investigated in Refs. 28) - 39).} 
In this paper, we consider an analytically 
continued two-parameter Melvin background which, before the Kaluza-Klein 
reduction, has the metric 
\begin{align}
\begin{aligned}
ds^2 = -dt^2 + \frac{t^2}{(1+q^2 t^2)(1 + p^2 t^2)}d\theta^2 + \f{1 + q^2 t^2}{1 + p^2 t^2}(dy + A_{\theta} 
d\theta)^2 ~,\\
B_{y \theta} = \f{p t^2}{1 + p^2 t^2} ~, ~~~~~
A_{\theta} = \frac{q t^2}{1+q^2 t^2} ~, ~~~~~
e^{2(\Phi - \Phi_0)} = \f{1}{1 + p^2 t^2} ~,
\label{KK2}
\end{aligned}
\end{align}
where we have $-\infty<t, \theta<\infty$, and $y\sim y+ 2\pi R$. 
Note that via the Wick rotations $t\rightarrow ir, \theta\rightarrow i\varphi, p\rightarrow ip$ and 
$q\rightarrow iq$, 
we can recover the usual two-parameter Euclidean Melvin metric.$^{42)}$ 
Now we see that the metric \eqref{KK2} reduces to an orbifold as follows. 
The sigma model associated with closed strings 
in the above background 
\eqref{KK2} is written as% 
\footnote{Though \eqref{KK2} is generated from one half of ${\mathbb R}^{1,3}$, 
it is possible to extend to the entire ${\mathbb R}^{1,3}$ using 
the same manipulation.}  
\begin{align}
S = \frac{1}{\pi \al}\int d^2 z \Big[ -\bp t \p t + \bp y \p y + \f{t^2}{1 +p^2 t^2}
\big(\bp {\theta} + (q - p)\bp y\big)
\big(\p \theta + (q + p)\p y\big) \Big] ~,
\label{tsig}
\end{align}
with the convention $\p = \f{1}{2}(\p_{\sigma} + 
\p_{\tau})$ and $\bar{\p} = \f{1}{2}(\p_{\sigma} - \p_{\tau})$. 
Closely following the treatment presented in 
Ref. 42), if one 
first compactifies $\theta$ to some finite radius and then 
T-dualizes it to $\tilde \theta$, 
next shifts $y$ as $y\rightarrow y'= y + b\tilde\theta$, subsequently 
T-dualizes 
$\tilde\theta$ to $\theta'$, and finally decompactifies $\theta$, 
the action $\eqref{tsig}$ becomes that of a free sigma model in terms of 
$x'^{\pm}=\f{1}{\sqrt{2}}t e^{\pm \theta'}$ and $y'$. Here, 
$(x'^{\pm},y')$ satisfies the 
periodicity conditions 
\begin{align}
x'^{\pm}(\sigma + 2\pi, \tau) = e^{\pm 2\pi\big(qRw + \al p(\f{n}{R} - q{J})\big)}x'^{\pm}(\sigma, \tau) ~,
&&y'(\sigma + 2\pi, \tau) = y'(\sigma, \tau) + 2\pi Rw - 2\pi \al p{J} ~.
\label{TsT}
\end{align}
Note that $(n,w)\in{\mathbb Z}$ and $J$ 
is the boost generator of the $x'^{\pm}$-plane. Another feature of the above action 
is that in the limit of vanishing $p$, 
$\eqref{TsT}$ reduces to the shifted-boost orbifold proposed in Ref. 3).

In Ref. 21), it is shown that 
the probe D-brane emits twisted 
closed strings in Misner space,%
\footnote{In static backgrounds, e.g. the 
near-horizon region of a stack of NS5-branes, closed string emission from D-branes can 
take place.$^{46)}$} which is the simplest time-dependent 
orbifold. 
Inspired by that, we set out to determine if there is 
similar behavior in our case when using 
the above free field representation to study the D-brane spectrum. 
We observe 
that the winding string coupling to D-branes is suggested by 
their classical profiles. On the other hand, 
by explicitly constructing 
their boundary states, this property becomes evident and 
we are also able to compute the winding string emission rate.

In addition, we note that the behavior of 
winding string emission can be traced to the 
open string Hamiltonian $H_o$ on these probe D-branes. 
This is because the open string annulus amplitude, 
\begin{align}
{\cal A}(t)=\int^\infty_0\f{dt}{t}\text{Tr}~e^{-2\pi tH_o} ~,
\end{align}
upon the moduli transformation $t\rightarrow s=\f{1}{t}$, 
has a form similar to the one-loop partition function 
${\cal A}_{\text{elec}}(t)$ of open strings subject to a 
constant electric field. 
As is widely known, in the latter case, 
worldsheet fields acquire twisted periodicity 
via a doubling trick, which is responsible for 
string pair creation.$^{47)}$ In our case, 
after moduli integration, ${\cal A}(s=\frac{1}{t})$ 
(the closed-channel cylinder amplitude) also 
acquires an imaginary part, and this, along with the optical theorem, 
accounts for the winding string emission.

The outline of this paper is as follows. 
In $\S$2, we examine the worldvolume theory of D-branes and 
construct their boundary states. Next, 
we explicitly compute the winding string emission rate. 
In the appendix, we calculate the torus amplitude. 
We find that when the B-field parameter $p$ is taken to zero, 
this amplitude is identical to 
that of the shifted-boost orbifold derived in Ref. 3).

\section{D-branes in the Lorentzian Melvin geometry}

In order to study the D-brane spectrum,%
\footnote{See Ref. 44) for the Euclidean counterparts.} 
we classify D-branes according to the free field representation 
$(x'^{\pm}=\f{1}{\sqrt{2}}t e^{\pm \theta'}, y')$ appearing in \eqref{TsT}. 
There are four kinds of boundary conditions we can impose on $(\theta', y')$, i.e. 
\begin{align}
\begin{aligned}
(\text{i}):~~ (N,N) ~,
&&(\text{ii}):~~ (N,D) ~,
&&(\text{iii}):~~ (D,D) ~,
&&(\text{iv}):~~ (D,N) ~,
\label{cl}
\end{aligned}
\end{align}
where $N$ ($D$) stands for the Neumann (Dirichlet) boundary condition. 
We restrict our attention to only 
those which carry the twisted sector, i.e. types $(\text{i})$ and $(\text{ii})$.%
\footnote{Note that the D-particle of type $(\text{iii})$ 
and its $y'$ direction T-dual counterpart, namely, that of type $(\text{ii})$, 
cannot couple to the twisted sector.}

\subsection{Classification of D-branes}

${\normalsize\textsc{Type (\text{i}): D2-branes}}$\\
In the following DBI analysis, the relevant coordinates are those of the curved 
metric \eqref{KK2}. 
The DBI action which 
describes the dynamics of slowly varying fields on a D2-brane is 
\begin{align}
S_{\text D2} = -\tau_{2}\int dt d{\theta}dy ~e^{-\Phi}\sqrt{-\det(G+B)} ~,
\label{DB}
\end{align}
where 
the D$p$-brane tension is expressed as 
$\tau_{p}= g^{-1}_s(2\pi)^{-p} \al^{-(p+1)/2}$. 
The energy-momentum tensor $\bf{T}^{\mu \nu}$ and the NS-source $\bf{S}^{\mu \nu}$ 
can be derived by infinitesimally varying the action w.r.t. 
$G^{\mu \nu}$ and $B^{\mu \nu}$ as 
\begin{align}
\f{\delta S_{\text D2}}{\delta (G+B)_{\mu \nu}} = -\f{\tau_{2}}{2}e^{-\Phi}\sqrt{-\det(G+B)}(G+B)^{\mu \nu} = ({\bf{T} + \bf{S}})^{\mu \nu} ~.
\end{align}
Plugging the metric $\eqref{KK2}$ into this relation 
(with $\mu,\nu=t,\theta,y$), we have 
\begin{eqnarray}
{\bf{T}}^{\mu \nu} = \f{-\tau_2 |t|}{2}
\left(
\begin{array}{ccc}
-1&0&0\\
0&\f{1+ q^2 t^2}{t^2}&-q\\
0&-q&1\\
\end{array}
\right) ~,~~~
{\bf{S}}^{\mu \nu} = \f{-\tau_2 |t|}{2}
\left(
\begin{array}{ccc}
0&0&0\\
0&0&p\\
0&-p&0\\
\end{array}
\right) ~.
\label{TS}
\end{eqnarray}
Because there appears a non-trivial worldvolume 
B-field, 
the non-commutative Yang-Mills theory on the D-brane is now 
described by a new set of $open~string$ parameters, namely, the 
open string metric ${\bf{G}}^{\mu \nu}$, the open string coupling ${\bf{G}_s}$, and 
the non-commutativity parameter $\Theta$. 
Making use of the 
Seiberg-Witten map, i.e. 
\begin{align}
(G+B)^{\mu \nu} = {\bf{G}}^{\mu \nu}+\f{\Theta^{\mu \nu}}{2\pi \al} ~,
&&{\bf{G}}_{\mu \nu} = (G-BG^{-1}B)_{\mu \nu} ~,
\end{align}
and 
\begin{align}
{\bf{G}_s}= e^{\Phi}g_s \sqrt{\frac{-\det{\bf{G}}}{-\det{(G+B)}}} ~,
\end{align}
we find 
 \begin{eqnarray}
{\bf{G}}_{\mu \nu} = 
\left(
\begin{array}{ccc}
-1&0&0\\
0&t^2&qt^2\\
0&qt^2&1+q^2 t^2\\
\end{array}
\right) ~,~~~~~ \Theta^{\theta y} = 2\pi\al p ~,~~~~~{\bf{G}_s}=g_s ~.
\label{om}
\end{eqnarray}
We can comment on the holographical dual of this open string theory for 
$l_s \rightarrow 0$ in large 't Hooft coupling.

Recall that the open (closed) 
string spectrum is truncated at 
massless modes when $l_s \rightarrow 0$, 
meanwhile bulk closed strings decouple from 
D-branes because the Newton constant $G_{10}\sim l_s^8 g_s^2$ goes to zero. 
Also, according to the 
usual AdS/CFT correspondence, the gauge theory on a stack of $N$ $(N\gg1)$ 
D3-branes is dual to the near-horizon geometry of 
a classical supergravity solution 
whose characteristic scale is about $l_s(g_s N)^{1/4}$. 
The near-horizon geometry of a stack of $N$ D3-branes 
extending over $(x^0,x^1,x^2,x^3)=(t\cosh\phi,t\sinh\phi,y,z)$ is 
\begin{align}
\f{ds^2}{\al}=\frac{U^2}{\sqrt{4\pi g_s N}} (-dt^2+t^2d\phi^2+dy^2+dz^2) 
+\sqrt{4\pi g_s N}(\frac{dU^2}{U^2}+d\Omega_5^2) ~.
\end{align}
Through the decoupling limit,$^{22),~ 23),~ 40),~ 41)}$ 
\begin{align}
\lim_{\al\rightarrow 0}\f{U}{\al}={\text {finite}}~,~~~~
\lim_{\al\rightarrow 0}\Theta^{\mu \nu}=2\pi\eta={\text {finite}} ~,~~~~
p=\f{\eta}{\al} ~,
\end{align}
the metric part of the gravity solution can thus be constructed as%
\footnote{We are grateful to A. Hashimoto for discussions regarding this point.}
\begin{align}
\f{ds^2}{\al}=\f{U^2}{\sqrt{\lambda}}(-dt^2 +dz^2)
+\f{\sqrt{\lambda}U^2}{\lambda+\eta^2 t^2 U^4}
\big(t^2(d\theta+qdy)^2 + dy^2 \big)
+\sqrt{\lambda}(\frac{dU^2}{U^2}+d\Omega_5^2) ~,
\label{sugra}
\end{align}
where $\lambda=g^2_{YM} N$ is the 't Hooft coupling. 
It is seen that the second term on the RHS of 
\eqref{sugra} differs from that in the flat case, 
due to the presence of a non-zero $p$ parameter. 
We leave the study of further implications 
of this gravity dual as a future work. 
\\

${\normalsize\textsc{type (\text{ii}): D1-branes}}$
\\
According to $\eqref{TsT}$, it is seen that the 
momentum conjugate to $y'$ is 
\begin{align}
\f{1}{2}(p'_L + p'_R) = 
(\f{n}{R} - q{J}) ~,~~~~~\f{1}{2}(p'_L - p'_R) = (\f{Rw}{\al} - p{J}) ~.
\label{LR}
\end{align}
This implies 
that the parameters $p$ and $q$ in \eqref{TsT} become exchanged 
through T-dualizing $y'$. Accordingly, 
the metric part in \eqref{KK2}, after T-dualizing $y'$, 
should be modified in the corresponding manner via $q\leftrightarrow p$ as 
\begin{align}
\begin{aligned}
ds^2=-dt^2+&\frac{1}{1 + q^2 t^2}( t^2 d\Theta^2 + dy^2) ~,~~
\Theta=(\theta + py) ~,~~
\left(
\begin{array}{c}
\Theta\\
y\\
\end{array}
\right) \sim
\left(
\begin{array}{c}
\Theta+2\pi pR'\\
y+2\pi R'
\end{array}
\right) ~,~~
R'=\frac{\al}{R} ~.
\label{22}
\end{aligned}
\end{align}
Note that the equations of motion of the sigma model on the metric $\eqref{22}$ 
are 
related to those of the aforementioned free 
field one as%
\footnote{We consider physics in the case of $t\ne0$.} 
\begin{align}
\begin{aligned}
(1 + q^2 t^2)\p_{\sigma}\theta' = \p_{\sigma}(\theta + py) + q\p_{\tau}y ~,\\
(1 + q^2 t^2)\p_{\tau}y' = \p_{\tau}y - qt^2\p_{\sigma}(\theta + py) ~,
\label{yu}
\end{aligned}
\end{align} 
which are obtained from the relation between two sigma models connected by 
T-duality.$^{43)}$ 
Based on this, the above type $(\text{ii})$ boundary condition 
can be transformed into 
\begin{align}
\begin{aligned}
\p_{\sigma}t = \p_{\sigma}(\theta +py)= \p_{\tau}y = 0 ~. 
\end{aligned}
\end{align}
Employing the static gauge ($\xi^0=t$, $\xi^1=\Theta$), where 
$\xi^{0}$ and $\xi^{1}$ parameterize the D-string 
worldvolume, we observe that its classical profile is $y=C$ ($C$: constant).

Due to the fundamental region $\mathbb{S}^1 \times \mathbb{R}$ (a cylinder), 
which comes from the identification in $\eqref{22}$, 
all images of this D-string, i.e. $y=C+2\pi m R'$ ($m\in{\mathbb Z}$), 
should be summed over. 
Hence, the D-string takes the form of a spiral that makes 
an angle $\vartheta=\arctan({1}/{pt})$ with $\mathbb{S}^1$ and couples to 
winding strings. 
In the case $\vartheta\rightarrow\f{\pi}{2}$, 
i.e. the background B-field parameter $p\rightarrow0$, 
the D-string becomes parallel to the cylinder axis. This suggests that 
winding mode coupling is no more possible. In other words, 
the non-vanishing coupling to winding modes can be understood as 
being 
induced by a non-zero $p$. 
This point is revisited below in the context of the boundary state.

\subsection{Boundary state}
To 
include the $\al$ corrections, we next study the 
boundary state formalism. For simplicity, 
we work out only the bosonic part, without loss of generality. 
Note that the 
relevant coordinates for the following analysis are those used in defining 
the free field representation \eqref{TsT}. 
Subsequently, these results are utilized to explicitly derive 
the winding string emission rate. \\

${\normalsize\textsc{Boundary state}}$
\\

We first consider the case of a D2-brane that 
wraps the entire $(x'^{\pm},y')$. From 
$\eqref{LR}$, the Neumann boundary condition along $y'$ 
implies 
\begin{align}
\begin{aligned}
\p_{\tau}y'|B\rangle\rangle_{\text {D2}}=0 &\rightarrow (p'_L + p'_R)|B\rangle\rangle_{\text {D2}} = 
(\f{n}{R}-q{J})|B\rangle\rangle_{\text {D2}} = 0 ~;
\label{ob}
\end{aligned}
\end{align}
that is, ${J}|B,n=0\rangle\rangle_{\text {D2}}=0$. 
Therefore, according to \eqref{TsT}, 
closed strings which couple to the D2-brane satisfy 
the periodicity condition 
\begin{align}
\begin{aligned}
x'^{\pm}(\tau,\sigma+2\pi)&=e^{\pm 2\pi \gamma w}x'^{\pm}(\tau,\sigma) ~,
&&\gamma=qR ~,
\end{aligned}
\end{align}
and the mode expansion of $x'^{\pm}$ reads 
\begin{align}
x'^{\pm} (\tau,\sigma) = 
i \sqrt{\frac{\alpha '}{2}}\sum_{m \in {\mathbb Z}}& 
\Big[\frac{\alpha^{\pm}_m}{m \pm i\gamma w} e^{-i(m \pm i\gamma w)(\tau + \sigma)} +
  \frac{\tilde \alpha^{\pm}_m}{m \mp i\gamma w}e^{-i(m \mp i\gamma w)
  (\tau - \sigma)}\Big] ~.
\end{align}
$\eqref{ob}$, together with the Neumann condition 
$\p_{\tau}x'^{\pm}|B\rangle \rangle_{{\text {D2}}} = 0$, 
dictates that the boundary state be of the form%
\footnote{The commutation relations are defined to be $[\alpha ^+_m,\alpha ^-_n]=(- m - i \nu )\delta_{m+n}, ~
[\tilde \alpha ^+_m, \tilde \alpha ^-_n]=(- m + i \nu )\delta_{m+n}$. 
For $\nu < 0$, $\alpha^+_0$ and $\tilde \alpha^-_0$ ($\alpha^-_0$ and 
$\tilde \alpha^+_0$) 
act as 
creation (annihilation) operators, and vice versa for $\nu > 0$.}
\begin{align}
 |B\rangle \rangle_{{\text {D2}}} = \sum_{w} {\cal{N}}_{w}
 \exp \left( \sum_{m \geq 1}
  \frac{\alpha^+_{-m} \tilde \alpha^-_{-m}}{m + i \gamma w}
   + \sum_{m \geq 0}
  \frac{\alpha^-_{-m} \tilde \alpha^+_{-m}}{m - i\gamma w} -\sum_{l\ge1}
  \f{\alpha^{y'}_{-l}\ti\alpha^{y'}_{-l}}{l} \right)|n=0, w\rangle
  \otimes|X_{\perp}\rangle\otimes|{\text {GH}}\rangle ~.
  \label{BD2}
 \end{align}
Here, the transverse components, $|X_{\perp}\rangle$, 
and the ghost part, $|{\text {GH}}\rangle$, are the 
usual flat space ones. 
To evaluate ${\cal{N}}_{w}$ explicitly, we compute the 
annulus amplitude as%
\footnote{Here, an overall volume factor is omitted 
and the untwisted sector is ignored.} 
\begin{align}
\begin{aligned}
Z(t) &= {\text {Tr}}~e^{{2\pi i\gamma w{J}}+{2\pi iRw p_{y'}}}
e^{-2\pi t L^{\text{flat}}_0} ~,\\
 \int_0^{\infty} \frac{dt}{t}Z(t) &=\int_0^{\infty} \frac{dt}{t} \sum_{w \neq 0}\frac{ie^{\f{-\pi^2 R^2 w^2}{2\pi\al t}}}
   {2\sinh (  \pi\gamma|w| ) (8 \pi^2 t \alpha ')^{\frac{1}{2}}
    \vartheta_1 (i\gamma |w| {\big|} it)\eta(it)^{21}} \\
&\rightarrow
     \int_0^{\infty} ds\sum_{w \neq 0}
    \frac{s^{\frac{-23}{2}} e^{- \pi \gamma^2 w^2 s-\frac{\pi R^2 w^2 s}{2 \alpha ' }}}
   {2 \sinh (  \pi\gamma |w| )
  (8 \pi^2 \alpha ')^{\frac{1}{2}}
   \vartheta_1 ( \gamma|w|s{\big|}is)\eta(is)^{21}} ~,
   \label{M}
   \end{aligned}
\end{align}
where the insertion in the first line reflects 
the twisted periodicity conditions. 
Also, we have carried out a moduli transformation in the third line. 
The Cardy condition then requires 
${\cal N}_w = \sqrt{\frac{8 \pi^2 \alpha'}{  2 \sinh (  \pi \gamma|w|) }}
{ \cal N}_{2}$, 
where ${\cal N}_{2}$ denotes the 
usual flat space normalization. Note that the $y'$ part of 
the above partition function can be recast into the form 
\begin{align}
Z^{\text{D2}}_{y'}(t) = {\text {Tr}}~e^{-2\pi t\al(\f{a}{R}-q{J})^2 + N_{y'} - \f{1}{24}} ~,&&
a\in{\mathbb Z} ~,
\label{to1}
\end{align}
which results from integrating out $p_{y'}$ 
and applying the Poisson resummation w.r.t. $w$.

Next, we construct the above D-string boundary state. 
In contrast to $\eqref{ob}$, $|B\rangle\rangle_{\text {D1}}$, which wraps
 only $x'^{\pm}$, satisfies 
\begin{align}
\p_{\sigma}y'|B\rangle\rangle_{\text {D1}}=0 \rightarrow (\f{Rw}{\al}-p{J})|B\rangle\rangle_{\text {D1}} = 0 ~,
\label{an2}
\end{align}
i.e. 
${J}|B,w=0\rangle\rangle_{\text {D1}}=0$. Together with 
$\p_{\tau}x'^{\pm}|B\rangle\rangle_{\text {D1}}=0$, we find that 
the twisted condition on closed strings coupled to the D-string is 
$x'^{\pm}(\sigma+2\pi)=e^{\pm 2\pi p\f{\al }{R}n}x'^{\pm}(\sigma)$. 
Based on these, the open string annulus amplitude 
is computed as 
\begin{align}
{\text {Tr}}~e^{{2\pi ip\f{\al}{R}n{J}}+{i\f{n}{R}y'}}
e^{-2\pi t L^{\text{flat}}_0} ~,
&&n\in {\mathbb Z} ~.
\label{ZY}
\end{align}
By integrating out $y'$ and applying the Poisson resummation w.r.t. $n$, 
it is straightforward to extract the $y'$ part of the partition function, 
\begin{align}
Z^{\text{D1}}_{y'}(t) = 
{\text {Tr}}~e^{-2\pi t\al(\f{mR}{\al}-p{J})^2 + N_{y'} - \f{1}{24}} ~.
\label{to}
\end{align}
The above expression elucidates the T-dual relation between $\eqref{to1}$ and $\eqref{to}$. 
The corresponding 
boundary state is of the form 
\begin{align}
\begin{aligned}
 | B\rangle \rangle_{{\text {D1}}} = \sum_{n}{\cal N}_n
 \exp &\left( \sum_{m \geq 1}
  \frac{\alpha^+_{-m} \tilde \alpha^-_{-m}}{m + i \xi n}
   + \sum_{m \geq 0}
  \frac{\alpha^-_{-m} \tilde \alpha^+_{-m}}{m - i \xi n} +
  \sum_{l\ge1}\f{\alpha^{y'}_{-l}\ti\alpha^{y'}_{-l}}{l} \right)|n, w=0\rangle
   \otimes|X_{\perp}\rangle\otimes|{\text {GH}}\rangle ~,
   \label{bd1}
  \end{aligned}
 \end{align}
with $\xi=p\f{\al}{R}$ and ${\cal N}_n=\sqrt{\frac{8\pi^2 \al}
{2\sinh (  \pi \xi|n|) }}{\cal N}_{1}$. \\

${\normalsize\textsc{Twisted string emission}}$
\\

Because the integrand in the third line of $\eqref{M}$ contains poles at $s=\f{l}{\gamma|w|}$, 
$l=1,2,3,\cdots$, 
an imaginary part can be gained after performing the moduli integral along 
a slightly deformed route over the complex plane. 
As asserted in Ref. 21), this imaginary part 
is responsible for the twisted string emission 
in accordance with the optical theorem in 
usual field theory. 
The emission rate is determined to be
\begin{align}
\begin{aligned}
-2{\text {Im}}&\Big[\frac{\alpha '\pi}{2}\int{ds}_{\text {D2}}\langle\langle B|
e^{- \pi s (L_0 + \tilde L_0)}|B\rangle\rangle_{\text {D2}}\Big]\\
  = &\sum_{w \neq 0}
   \sum_{l=1}^{\infty} \frac{(-1)^{l+1}}{2 \g|w| \sinh (\pi \g|w|) (8 \pi ^2 \alpha ')^{\frac{1}{2}}}
    \left( \frac{\g|w|}{l} \right)^{\frac{23}{2}}
   \sum_{{\text {states}}}  
   e^{- \frac{2 \pi l}{\g|w|} (\frac{R^2 w^2}{4 \alpha '} + N - 1 ) - \pi l \g|w|} ~.
     \end{aligned}
\end{align}
Correspondingly, that of the codimension-one D-string is 
\begin{align}
\begin{aligned}
-2{\text {Im}}&\Big[
\frac{\alpha '\pi}{2}
  \int {ds}_{\text {D1}}\langle\langle B|
 e^{- \pi s (L_0 + \tilde L_0 )}
|B\rangle\rangle_{\text {D1}}\Big] \\
  = &\f{{\cal N}^2_1}{{\cal N}_2^2}\sum_{n \neq 0}
   \sum_{l=1}^{\infty} \frac{(-1)^{l+1}}{2 \xi|n| \sinh (\pi \xi|n|) 
   (8 \pi ^2 \alpha ')^{\frac{1}{2}}}
    \left( \frac{\xi|n|}{l} \right)^{\frac{23}{2}}
     \sum_{\text {states}} e^{- \frac{2 \pi l}{\xi|n|} 
(\frac{\al n^2}{4 R^2} + {N} - 1 ) - \pi l \xi|n|} ~,
\end{aligned}
\end{align}
where the prefactor is ${{\cal N}^2_1}/{{\cal N}_2^2}=4\pi^2 \al$.

The reason that the D-string sources winding strings can be traced to the 
open string Hamiltonian in $\eqref{to}$, where the term $(\f{mR}{\al}-p{J})^2$ 
rules this key dynamics. In summary, the mechanism closely resembles that of 
open string pair creation in a constant electric field. 
Let us investigate this point.

In fact, due to a non-zero $p$, we see that 
a projection in \eqref{ZY} is encoded, i.e. only states invariant under $e^{2\pi i{J}}$ 
survive after traveling around a loop in spacetime. 
Using the worldsheet open-closed duality to exchange $(\tau,\sigma)$, 
we can treat this projection as a twisted periodicity condition 
imposed on worldsheet fields. 
This situation strongly resembles the case of 
open strings subject to a constant electric field $E$. In the latter case, 
worldsheet fields acquire twisted periodicity through a doubling trick. 
That is, 
the lightcone coordinates of open strings satisfy the relation 
$x^{\pm}(\sigma+2\pi)=e^{\pm2\pi\beta}x^{\pm}(\sigma)$, where 
$E=\tanh\pi\beta>0$, as is shown in Ref. 47). 
Hence, the 
one-loop partition function for the $x^{\pm}$ part reads 
\begin{align}
Z^{\pm}_{ E}(s')\sim\frac{ \eta (is')}{\vartheta_1 (\beta s'|is')}
   e^{ - \pi s'\beta^2} ~,
\label{E}
\end{align}
where $s'$ is the annulus modulus. 
In our case, the moduli-transformed annulus amplitude, for example, in 
$\eqref{M}$ is 
\begin{align}
Z_{\text{closed}}^{\pm}(s)\sim
\f{\eta(is)e^{-\pi s\nu^2}}{2\sinh (\pi\nu)\vartheta_1 (s\nu|is)}  ~,
\label{mod}
\end{align}
where $s$ is the cylinder modulus, and 
$\nu$ is the twist parameter. 
Just as in the case of \eqref{E}, which, 
after the moduli integral, leads to 
open string pair creation,$^{47)}$ 
$Z_{\text{closed}}^{\pm}(s)$, 
together with the optical theorem, 
accounts for the winding string 
emission, 
after $s$ is integrated out.

\section*{Acknowledgements}

I am grateful to Y. Matsuo, T. Takayanagi, 
A. Hashimoto, J. Raeymaekers, Y. Tachikawa, Y. Imamura, Y. Nakayama 
and, especially, Y. Hikida for useful discussions.

%%%%%%%%%%%%%%%%%%%%%%%%%%%%%%%%%%%%%%%%%%%%%%%%%%%%%%%%%%%%%%%%%%%%%

\appendix
\section{Torus amplitude} 

Here, we compute 
the closed string torus amplitude based on \eqref{TsT}. 
Note that the orbifold \eqref{TsT} can be deformed smoothly to a 
shifted-boost orbifold by 
taking $p\to0$. It is non-trivial to 
determine whether, as all oscillating modes are considered, the one-loop 
partition function is identical to that of the shifted-boost orbifold first 
computed in Ref. 3).

Adding seven flat spectator directions, we embed the Lorentzian Melvin 
geometry into 10-dimensional superstring theory. 
The partition function in the operator formalism can be written as 
\begin{align}
\begin{aligned}
 Z = &\big[{{\rm Tr}_{\text {NS}}} {\textstyle \frac12}(1 + (-1)^F)q^{L_0} -
      {{\rm Tr}_{\text {R}}}{\textstyle \frac12}(1 + (-1)^F)q^{L_0}\big]\\
      \times
         &\big[{{\rm Tr}_{\text {NS}}} {\textstyle \frac12}(1 + (-1)^{\ti F})\bar{q}^{\ti L_0} -
      {{\rm Tr}_{\text {R}}}{\textstyle \frac12}(1 \mp (-1)^{\ti F})
      \bar{q}^{\ti L_0}\big] ~,
\label{tor}
\end{aligned}
\end{align}
where $q = e^{2\pi i\tau}$, $
\tau = \tau_1 + i\tau_2$, and $-$ $(+)$ is assigned to the type IIA (IIB) theory. 
In terms of the free field 
representation $(x'^{\pm},y')$, the Virasoro generators are 
\begin{align}
\begin{aligned}
L_0 &= c
 + \f{\al}{4}p^{' 2}_L + \f{\al}{4}\vec{p}_7^2 + \nu{J}_L
      +  N ~, \\
 \tilde L_0 &= \ti c 
 + \f{\al}{4}p^{' 2}_R + \f{\al}{4}\vec{p}_7^2 - \nu{J}_R
 + \ti N ~,
\end{aligned}
\end{align}
where $\nu = qRw + p\al(\f{n}{R} - q{J})$, 
and $c$, $\ti c$ are the usual untwisted ordering constants.

Inserting a delta function into $\eqref{tor}$, as in Ref. 45), namely, 
\begin{align}
\begin{aligned}
1=\int d^2{j} ~\delta({{J}}_L - j)\delta({{J}}_R - \bar{j}) ~
;~~\delta({{J}}_L - j)\delta({{J}}_R - \bar{j})=
\int d^2\chi ~e^{2\pi i\chi ({{J}}_L - j)
 +2\pi i\bar{\chi}({{J}}_R - \bar{j})} ~,
\label{del}
\end{aligned}
\end{align}
we can rewrite the partition function as $(J=j+\bar j)$
\begin{align}
\begin{aligned}
Z(\tau)_{\text {super}} &= \f{V_{7}}{(2\pi)^{7}}(\al\tau_2)^{\f{-7}{2}}
\sum_{n, w\in{\mathbb Z}} \int d^2\chi d^2 j
\Big[{\text{Tr}} ~e^{2\pi i\chi{{J}}_L} q^{{N} -c}\Big]
\Big[{\text {Tr}} ~e^{2\pi i\bar \chi{J}_R} \bar q^{\ti{N} -\ti c}\Big] \\
&\times e^{2\pi i \tau_1(nw - \nu{{J}})} e^{-\pi \tau_2\al\big((\f{n}{R} - q{J})^2 + (\f{Rw}{\al} - p{J})^2\big)}
\exp{\Big[-2\pi i\chi j - 2\pi i\bar\chi \bar j + 2\pi i\tau\nu j
 + 2\pi i\bar{\tau}\nu\bar j\Big]} ~.
\label{Z}
\end{aligned}
\end{align}
To deal with the second line, we apply the 
Poisson resummation%
\footnote{$\sum_n \exp(-\pi a n^2 + 2\pi i b n)=
\f{1}{\sqrt{a}}\sum_m\exp\big(\f{-\pi(m-b)^2}{a}\big)$} 
w.r.t. $n$ and obtain 
\begin{align}
\begin{aligned}
R(\al\tau_2)^{\f{-1}{2}}\sum_{m, w\in{\mathbb Z}}&\exp
\Big[\f{-\pi R^2}{\al\tau_2}\big(K+2i\mu\tau_2 \bar j\big)
\big(\bar{K}-2i\mu\tau_2 j\big)\Big]\\
\times &e^{-2\pi iqR(m-\tau_1 w){J} - 2\pi qR\tau_2 w
(j-\bar j)}e^{-2\pi i\chi j - 2\pi i\bar\chi \bar j} ~,
\label{i}
\end{aligned}
\end{align}
where $K = m-\tau w$ and $\mu=\f{\al p}{R}$. Then, by inserting the identity
\begin{align}
1 = \f{1}{\al\tau_2}\int d\Xi d\bar{\Xi} \exp\big[\f{-\pi}{\al\tau_2}
(\Xi+RK+2i{\al}p\tau_2 \bar j)(\bar{\Xi}-R\bar{K}+2i{\al}p\tau_2 j)\big] ~,
\end{align}
$\eqref{i}$ becomes 
\begin{align}
R(\al\tau_2)^{\f{-3}{2}}\int d\Xi d\bar{\Xi}\sum_{m, w\in{\mathbb Z}}
\exp\big[\f{-\pi}{\al\tau_2}(\Xi\bar{\Xi} - \Xi R\bar{K} + \bar{\Xi}RK)
- 2\pi i(\chi+\Theta)j - 2\pi i(\bar\chi+\bar{\Theta})\bar j\big] ~,
\end{align}
where $\Theta = p\Xi + qRK$. 
Finally, evaluating the trace in $\eqref{Z}$ and 
integrating out $j,\bar j,\chi$ and $\bar\chi$, 
we obtain%
\footnote{That no negative norm sates propagate in this kind 
of Lorentzian torus amplitude is proven in Ref. 21).}
\begin{align}
\begin{aligned}
Z(\tau)_{\text {super}} &= \f{V_{7}R}{(2\pi)^{7}}(\al\tau_2)^{-5}\sum_{m, w\in{\mathbb Z}}\int d\Xi d\bar{\Xi}
\f{|\vartheta_3 (\Theta|\tau) \vartheta_3 (0|\tau)^3
        - \vartheta_4 (\Theta|\tau) \vartheta_4 (0|\tau)^3
        - \vartheta_2 (\Theta|\tau) \vartheta_2 (0|\tau)^3|^2}{4{|\eta(\tau)|^{18}}{|\vartheta_1 (\Theta|\tau)|^2}}\\
&\times
\exp\big[\f{-\pi}{\al\tau_2}(\Xi\bar{\Xi} - \Xi R\bar{K} + \bar{\Xi}RK)\big] ~.
\label{spo}
\end{aligned}
\end{align}
Next, using in $\eqref{spo}$ the following identity 
\begin{align}
\begin{aligned}
|\vartheta_3 (\Theta|\tau) \vartheta_3 (0|\tau)^3
        - \vartheta_4 (\Theta|\tau) \vartheta_4 (0|\tau)^3
        - \vartheta_2 (\Theta|\tau) \vartheta_2 (0|\tau)^3|^2 = 4|\vartheta_1 (\f{\Theta}
	{2}{\big|}\tau)^4|^2 ~,
\end{aligned}
\end{align}
we know that the supersymmetry is broken. 
Further, the modular 
invariance of $\int\f{d^2\tau}{\tau_2}Z(\tau)$ can also be seen 
under the transformation 
$\tau\rightarrow\f{-1}{\tau}$, $(m, w)\rightarrow(w, -m)$ and 
$(\Xi, \bar{\Xi})\rightarrow(\f{\Xi}{\tau}, \f{\bar{\Xi}}{\bar{\tau}})$.

By taking the limit of vanishing B-field ($p\rightarrow0$), 
the bosonic part of $\eqref{spo}$ involving only $(x'^{\pm}, y')$ 
is extracted to be
\begin{align}
\begin{aligned}
\lim_{p\rightarrow0}Z_{3} = R(\al\tau_2)^\f{-1}{2}\sum_{m, w\in{\mathbb Z}}\exp(\f{-\pi R^2 K\bar{K}}{\al\tau_2}-2\pi\tau_2
q^2R^2w^2)~\big|\vartheta_1(iqRK|\tau)\big|^{-2} ~.
\label{Z3}
\end{aligned}
\end{align}
The result is just identical to that of the shifted-boost 
orbifold previously derived in Ref. 3).

%%%%%%%%%%%%%%%%%%%%%%%%%%%%%%%%%%%%%%%%%%%%%%%%%%%%%%%%%%%%%%%%%%%%%%%%%
%%%%%%%%%%%%%%%%%%%%%%%%%%%%%%%%%%%%%%%%%%%%%%%%%%%%%%%%%%%%%%%%%%%%%%%%%


\begin{thebibliography}{99}

\bibitem{KOSST}
  J.~Khoury, B.~A.~Ovrut, N.~Seiberg, P.~J.~Steinhardt and N.~Turok,
   ``From big crunch to big bang,''
  Phys.\ Rev.\ D {\bf 65}, 086007 (2002)
  [arXiv:hep-th/0108187].
  %%CITATION = HEP-TH 0108187;%%

\bibitem{BHKN}
  V.~Balasubramanian, S.~F.~Hassan, E.~Keski-Vakkuri and A.~Naqvi,
  ``A space-time orbifold: A toy model for a cosmological singularity,''
  Phys.\ Rev.\ D {\bf 67}, 026003 (2003)
  [arXiv:hep-th/0202187].
  %%CITATION = HEP-TH 0202187;%%

\bibitem{CC}
  L.~Cornalba and M.~S.~Costa,
  ``A new cosmological scenario in string theory,''
  Phys.\ Rev.\ D {\bf 66}, 066001 (2002)
  [arXiv:hep-th/0203031].
  %%CITATION = HEP-TH 0203031;%%

\bibitem{Nekrasov}
  N.~A.~Nekrasov,
  ``Milne universe, tachyons, and quantum group,''
  Surveys High Energ.\ Phys.\  {\bf 17}, 115 (2002)
  [arXiv:hep-th/0203112].
  %%CITATION = HEP-TH 0203112;%%

\bibitem{Simon}
  J.~Simon, ``The geometry of null rotation identifications,'' JHEP
  {\bf 0206}, 001 (2002) [arXiv:hep-th/0203201].
  %%CITATION = HEP-TH 0203201;%%

\bibitem{LMS1}
  H.~Liu, G.~W.~Moore and N.~Seiberg,
  ``Strings in a time-dependent orbifold,''
  JHEP {\bf 0206}, 045 (2002)
  [arXiv:hep-th/0204168].
  ``Strings in time-dependent orbifolds,''
  JHEP {\bf 0210}, 031 (2002)
  [arXiv:hep-th/0206182].
  %%CITATION = HEP-TH 0206182;%%
  %%CITATION = HEP-TH 0204168;%%

\bibitem{CCK} L. Cornalba, M. S. Costa, and C. Kounnas,
  ``A resolution of the cosmological singularity with
  orientifolds'', Nucl. \ Phys. \ B{\bf 637} 378 (2002) 
  [arXiv:  hep-th/0204261.]
  %%CITATION = HEP-TH 0204261%%

\bibitem{Lawrence}
  A.~Lawrence, ``On the instability of 3D null singularities,'' JHEP
  {\bf 0211}, 019 (2002) [arXiv:hep-th/0205288].
  %%CITATION = HEP-TH 0205288;%%

\bibitem{FM}
  M.~Fabinger and J.~McGreevy, ``On smooth time-dependent orbifolds
  and null singularities,'' JHEP {\bf 0306}, 042 (2003)
  [arXiv:hep-th/0206196].
  %%CITATION = HEP-TH 0206196;%%

\bibitem{HP}
  G.~T.~Horowitz and J.~Polchinski, ``Instability of space-like and
  null orbifold singularities,'' Phys.\ Rev.\ D {\bf 66}, 103512
  (2002) [arXiv:hep-th/0206228].
  %%CITATION = HEP-TH 0206228;%%
  
\bibitem{NW}
  C.~R.~Nappi and E.~Witten,
   ``A closed, expanding universe in string theory,''
  Phys.\ Lett.\ B {\bf 293}, 309 (1992)
  [arXiv:hep-th/9206078].
  %%CITATION = HEP-TH 9206078;%%

\bibitem{EGKR}
  S.~Elitzur, A.~Giveon, D.~Kutasov and E.~Rabinovici,
   ``From big bang to big crunch and beyond,''
  JHEP {\bf 0206}, 017 (2002)
  [arXiv:hep-th/0204189].
  %%CITATION = HEP-TH 0204189;%%

\bibitem{CKR}
  B.~Craps, D.~Kutasov and G.~Rajesh,
   ``String propagation in the presence of cosmological singularities,''
  JHEP {\bf 0206}, 053 (2002)
  [arXiv:hep-th/0205101].
  %%CITATION = HEP-TH 0205101;%%

\bibitem{HT}
  Y.~Hikida and T.~Takayanagi,
   ``On solvable time-dependent model and rolling closed string tachyon,''
  Phys.\ Rev.\ D {\bf 70}, 126013 (2004)
  [arXiv:hep-th/0408124].
  %%CITATION = HEP-TH 0408124;%%

\bibitem{TT}
  N.~Toumbas and J.~Troost,
   ``A time-dependent brane in a cosmological background,''
  JHEP {\bf 0411}, 032 (2004)
  [arXiv:hep-th/0410007].
  %%CITATION = HEP-TH 0410007;%%

\bibitem{ohta2}
  T.~Ishino, H.~Kodama and N.~Ohta,
 ``Time-dependent Solutions with Null Killing Spinor in M-theory and Superstrings,''
Phys.\ Lett.\ B {\bf 631}, 68 (2005)
[arXiv:hep-th/0509173].
  %%CITATION = HEP-TH 0509173;%%

\bibitem{STV}
  M.~Spradlin, T.~Takayanagi and A.~Volovich,
  ``String theory in $\beta$ deformed spacetimes,''
  JHEP {\bf 0511}, 039 (2005)
  [arXiv:hep-th/0509036.]
  %%CITATION = HEP-TH 0509036;%%
  
\bibitem{BCKR}
  M.~Berkooz, B.~Craps, D.~Kutasov and G.~Rajesh,
   ``Comments on cosmological singularities in string theory,''
  JHEP {\bf 0303}, 031 (2003)
  [arXiv:hep-th/0212215].
  %%CITATION = HEP-TH 0212215;%%
  
\bibitem{Pioline1}
  M.~Berkooz and B.~Pioline, ``Strings in an electric field, and the
  Milne universe,'' JCAP {\bf 0311}, 007 (2003)
  [arXiv:hep-th/0307280], 
  M.~Berkooz, B.~Pioline and M.~Rozali, ``Closed strings in Misner
  space: Cosmological production of winding strings,''
  JCAP {\bf 0408}, 004 (2004)
  [arXiv:hep-th/0405126], B.~Durin and B.~Pioline,
  ``Closed strings in Misner space: A toy model for a big bounce?,''
  arXiv:hep-th/0501145.

\bibitem{Pioline3}
  M.~Berkooz, B.~Durin, B.~Pioline and D.~Reichmann,
  ``Closed strings in Misner space: Stringy fuzziness with a twist,''
  JCAP {\bf 0410}, 002 (2004)
  [arXiv:hep-th/0407216].
  %%CITATION = HEP-TH 0407216;%%




\bibitem{HNP2}
  Y.~Hikida, R.~R.~Nayak and K.~L.~Panigrahi,
  ``D-branes in a big bang/big crunch universe: Misner space,''
  JHEP {\bf 0509}, 023 (2005)
  [arXiv:hep-th/0508003.]
  %%CITATION = HEP-TH 0508003;%%

\bibitem{HS}
  A.~Hashimoto and S.~Sethi,
   ``Holography and string dynamics in time-dependent backgrounds,''
  Phys.\ Rev.\ Lett.\  {\bf 89}, 261601 (2002)
  [arXiv:hep-th/0208126].
  %%CITATION = HEP-TH 0208126;%%

\bibitem{AP}
  M.~Alishahiha and S.~Parvizi,
   ``Branes in time-dependent backgrounds and AdS/CFT correspondence,''
  JHEP {\bf 0210}, 047 (2002)
  [arXiv:hep-th/0208187].
  %%CITATION = HEP-TH 0208187;%%

\bibitem{DN}
  L.~Dolan and C.~R.~Nappi,
   ``Noncommutativity in a time-dependent background,''
  Phys.\ Lett.\ B {\bf 551}, 369 (2003)
  [arXiv:hep-th/0210030].
  %%CITATION = HEP-TH 0210030;%%

\bibitem{CLO}
  R.~G.~Cai, J.~X.~Lu and N.~Ohta,
  ``NCOS and D-branes in time-dependent backgrounds,''
  Phys.\ Lett.\ B {\bf 551}, 178 (2003)
  [arXiv:hep-th/0210206].
  %%CITATION = HEP-TH 0210206;%%

\bibitem{Okuyama}
  K.~Okuyama,
  ``D-branes on the null-brane,''
  JHEP {\bf 0302}, 043 (2003)
  [arXiv:hep-th/0211218].
  %%CITATION = HEP-TH 0211218;%%

\bibitem{HNP}
  Y.~Hikida, R.~R.~Nayak and K.~L.~Panigrahi,
  ``D-branes in a big bang/big crunch universe: Nappi-Witten gauged WZW
  model,''
  JHEP {\bf 0505}, 018 (2005)
  [arXiv:hep-th/0503148].
  %%CITATION = HEP-TH 0503148;%%

\bibitem{Hagedorn}
  M.~S.~Costa, C.~A.~R.~Herdeiro, J.~Penedones and N.~Sousa,
  ``Hagedorn transition and chronology protection in string theory,''
  arXiv:hep-th/0504102.
  %%CITATION = HEP-TH 0504102;%%


\bibitem{MS}
  J.~McGreevy and E.~Silverstein,
  ``The tachyon at the end of the universe,''
  arXiv:hep-th/0506130.
  %%CITATION = HEP-TH 0506130;%%

\bibitem{Silverstein}
  E.~Silverstein,
  ``Dimensional mutation and spacelike singularities,''
  arXiv:hep-th/0510044.
  %%CITATION = HEP-TH 0510044;%%

\bibitem{APS}
  Y.~Hikida. and T.-S.~Tai,
  ``D-instantons and Closed String Tachyons in Misner Space,''
   [arXiv:hep-th/0510129].
  %%CITATION = HEP-TH 0510129;%%
  
  \bibitem{BKRS}
  M.~Berkooz, Z.~Komargodski, D.~Reichmann and V.~Shpitalnik,
  ``Flow of geometries and instantons on the null orbifold,''
  arXiv:hep-th/0507067.
  %%CITATION = HEP-TH 0507067;%%


\bibitem{CSV}
  B.~Craps, S.~Sethi and E.~P.~Verlinde,
  ``A matrix big bang,''
  arXiv:hep-th/0506180.
  %%CITATION = HEP-TH 0506180;%%

\bibitem{Li}
  M.~Li,
  ``A class of cosmological matrix models,''
  arXiv:hep-th/0506260.
  %%CITATION = HEP-TH 0506260;%%

\bibitem{LS}
  M.~Li and W.~Song,
  ``Shock waves and cosmological matrix models,''
  arXiv:hep-th/0507185.
  %%CITATION = HEP-TH 0507185;%%

\bibitem{She}
  J.~H.~She,
  ``A matrix model for Misner universe,''
  arXiv:hep-th/0509067. 
  ``Winding String Condensation and Noncommutative Deformation of Spacelike Singularity,''
arXiv:hep-th/0512299.
  %%CITATION = HEP-TH 0512299;%%
  %%CITATION = HEP-TH 0509067;%%

\bibitem{pp}
  S.~R.~Das and J.~Michelson,
  ``pp wave big bangs: Matrix strings and shrinking fuzzy spheres,''
  arXiv:hep-th/0508068.
  %%CITATION = HEP-TH 0508068;%%

\bibitem{Chen}
  B.~Chen,
  ``The time-dependent supersymmetric configurations in M-theory and matrix
  models,''
  arXiv:hep-th/0508191.
  %%CITATION = HEP-TH 0508191;%%

\bibitem{RS}
  D.~Robbins and S.~Sethi,
  ``A matrix model for the null-brane,''
  arXiv:hep-th/0509204.
  %%CITATION = HEP-TH 0509204;%%
  
 
  \bibitem{H}
  A.~Hashimoto and K.~Thomas,
  ``Non-commutative gauge theory on D-branes in Melvin Universes,''
  arXiv:hep-th/0511197.
  %%CITATION = HEP-TH 0511197;%%
  
  \bibitem{A}
  M.~Alishahiha, B.~Safarzadeh and H.~Yavartanoo,
  ``On Supergravity Solutions of Branes in Melvin Universes,''
  arXiv:hep-th/0512036.
  %%CITATION = HEP-TH 0512036;%%

\bibitem{TM1}
  T.~Takayanagi, T.~Uesugi,
  ``Orbifolds as Melvin Geometry,''
  JHEP {\bf 0112}, 004 (2001) 
  [arXiv:hep-th/0110099].
  %%CITATION = HEP-TH 0110099;%%
  


\bibitem{TM3}
  T.~Takayanagi, T.~Uesugi,
  ``Flux Stabilization of D-branes in NSNS Melvin Background,''
Phys.\ Lett.\ {\bf B528}, 156 (2002) 
  [arXiv:hep-th/0112199].
  %%CITATION = HEP-TH 0112199;%%

  
\bibitem{TM2}
  T.~Takayanagi, T.~Uesugi,
  ``D-branes in Melvin Background,''
  JHEP {\bf 0111}, 036 (2001) 
  [arXiv:hep-th/0110200], 
  E.~Dudas, J.~Mourad,
  ``D-branes in String theory Melvin backgrounds,''
  Nucl. Phys. {\bf {B 622}}, 46 (2002) 
  [arXiv:hep-th/0110186].
  
  
\bibitem{H2}
  A.~Hashimoto and L.~P.~Zayas,
  ``Correspondence Principle for Black Holes in Plane Waves,''
JHEP {\bf 0403} 014 (2004)
  [arXiv:hep-th/0401197].
  %%CITATION = HEP-TH 0401197;%%
  
  
\bibitem{Kutasov}
  D.~Kutasov,
  ``D-brane dynamics near NS5-branes,''
  arXiv:hep-th/0405058. Y.~Nakayama, Y.~Sugawara and H.~Takayanagi,
  ``Boundary states for the rolling D-branes in NS5 background,''
  JHEP {\bf 0407}, 020 (2004)
  [arXiv:hep-th/0406173]. D.~A.~Sahakyan,
  ``Comments on D-brane dynamics near NS5-branes,''
  JHEP {\bf 0410}, 008 (2004)
  [arXiv:hep-th/0408070]. B.~Chen, M.~Li and B.~Sun,
  ``D-brane near NS5-branes: With electromagnetic field,''
  JHEP {\bf 0412}, 057 (2004)
  [arXiv:hep-th/0412022]. Y.~Nakayama, K.~L.~Panigrahi, S.-J.~Rey and H.~Takayanagi,
  ``Rolling down the throat in NS5-brane background: The case of electrified
  D-brane,''JHEP {\bf 0501}, 052 (2005)
  [arXiv:hep-th/0412038]. Y.~Nakayama, S.-J.~Rey and Y.~Sugawara,
  ``D-brane propagation in two-dimensional black hole geometries,''
  arXiv:hep-th/0507040.

\bibitem{BP}
  C.~Bachas and M.~Porrati,
  ``Pair creation of open strings in an electric field,''
  Phys.\ Lett.\ B {\bf 296}, 77 (1992)
  [arXiv:hep-th/9209032]. C.~Bachas,
  ``D-brane dynamics,''
  Phys.\ Lett.\ B {\bf 374}, 37 (1996)
  [arXiv:hep-th/9511043].
  %%CITATION = HEP-TH 9209032;%%
  %%CITATION = HEP-TH 9511043;%%
  

\end{thebibliography}
\end{document}